\documentclass[conference,letterpaper]{IEEEtran}
 
\usepackage{amssymb}
\usepackage{graphicx}
\usepackage[utf8]{inputenc} 
\usepackage[T1]{fontenc}
\usepackage{url}
\usepackage{ifthen}
\usepackage{cite}
\usepackage[cmex10]{amsmath} 
\usepackage{amsmath,amsthm}

\usepackage{algorithm}
\usepackage{algpseudocode}
\newtheorem{thm}{Theorem}[section]

\newtheorem{example}[thm]{Example}
\newtheorem*{example*}{Example}

\newtheorem*{remark*}{Remark}
\newtheorem{proposition}[thm]{Proposition}

\begin{document}
\title{Outer Code Designs for Augmented and Local-Global Polar Code Architectures
} 

\author{%
  \IEEEauthorblockN{Ziyuan Zhu and Paul H. Siegel}
  \IEEEauthorblockA{
                    Department of Electrical and Computer Engineering, CMRR, University of California, San Diego\\
                    Email: \{ziz050, psiegel\}@ucsd.edu}
}

\maketitle

\begin{abstract}
In this paper, we introduce two novel methods to design outer polar codes for two previously proposed concatenated polar code architectures: augmented polar codes and local-global polar codes. These methods include  a stopping set (SS) construction and a nonstationary density evolution (NDE) construction. 
Simulation results demonstrate the advantage of these methods over previously proposed constructions based on density evolution (DE) and LLR evolution. 


\end{abstract}

\section{Introduction}

Polar codes, introduced by E. Ar\i kan~\cite{Arikan2009}, occupy a unique place in the history of error correction codes as the first family of codes to achieve Shannon capacity of arbitrary binary  symmetric  memoryless channels (BSMs). Ar\i kan~\cite{Arikan2011} also introduced the concept of systematic polar encoding, achieved through the solution of linear encoding equations that ensures the codewords contain the information bits at designated positions. Concatenated polar codes have been proposed that leverage the error floor performance of polar codes in conjunction with other powerful codes such as Low-Density Parity-Check codes~\cite{Eslami2013} and Reed-Solomon codes~\cite{Bakshi2010}. Expanding upon~\cite{Guo2014}, Elkelesh et al.~\cite{Elk2017} introduced an augmented polar code architecture that concatenates two polar codes, using an outer auxiliary polar code to further protect the semipolarized bit-channels within the inner polar code. In the same work, they also suggested connecting several inner polar codes through a single auxiliary polar code, offering the flexibility of codeword lengths other than a power of two. Motivated by practical applications in data storage and low-latency communication systems, Zhu et al.~\cite{Zhu2022} proposed an architecture for polar codes offering local-global decoding. In this scheme, a codeword comprising several inner polar codes is concatenated with a systematic outer polar code, thus enabling both local decoding of  the inner codes and global decoding of the codeword.

The belief propagation (BP) decoder for polar codes was introduced to increase throughput through a pipelined decoding process~\cite{ArikanBP}. While the BP decoder surpasses the error rate performance of the original successive-cancellation (SC) decoder, it still falls short of the SC-List (SCL) decoder~\cite{SCL}. The BP-List (BPL) decoder~\cite{BPL}, which incorporates different permutation patterns of BP decoding units, significantly enhances error rate performance, bridging the performance gap between BP-based and SC-based decoders. 

Polar codes and Reed-Muller (RM) codes share the same basic encoding matrix before selecting the information set: 
RM codes select rows according to their Hamming weights, while polar codes select rows by comparing their associated Bhattacharyya parameters~\cite{Arikan2009}. Another frozen set selection method, introduced by Mori et al.~\cite{DE}, uses density evolution (DE) to analyze  BP results for each decoding tree corresponding to the SC decoding process. The high computational complexity of DE motivated the Gaussian approximation (GA) algorithm~\cite{GA},  which assumes that the log-likelihood ratio (LLR) distribution corresponding to each variable node is a Gaussian with mean $m$ and variance $\sigma^2=2m$, thus reducing the convolution of densities to a one-dimensional computation of mean values. 
In~\cite{GA Polar}, Dai et al. proposed a modification to GA to address
the performance loss incurred when applying GA to long polar codes.

While design methods based on the Bhattacharyya parameter~\cite{Arikan2009}, DE, and GA were originally used in the context of SC decoding, they have also been applied to code design for BP decoding. Eslami et al.~\cite{Eslami2013} introduced a construction method based on stopping sets in the sparse polar code factor graph, aimed at increasing the stopping distance of the polar code. They provided empirical evidence showing improved performance under BP decoding, compared with the conventional code design. Another design approach based on  LLR evolution was proposed by Qin et al.~\cite{llr evolution}. In this method, weak bit-channels, identified  using LLR distributions obtained from BP decoder simulations, are swapped with stronger ones. While these algorithms have been shown to improve polar code design for BP decoding, the quest for an optimal design method remains an ongoing exploration.

In this paper, we propose two construction methods for designing outer codes within concatenated polar code architectures under BP decoding. The first construction, stopping set (SS) design, uses an analysis of stopping sets in the concatenated factor graph to identify the information set of the outer code. The nonstationary density evolution (NDE), initializes the DE construction of the outer code with empirical LLR densities from BP decoding of the inner code(s) to better reflect bit-channel reliabilities. Error rate simulations demonstrate that both of these methods can improve the performance of augmented and local-global polar codes. 


The paper is organized as follows. Section II briefly reviews background results and notations used in the rest of the paper. In Section III, the SS and NDE construction methods are presented and discussed. Section IV provides error rate  simulation results, and Section V concludes the paper.

\section{Preliminaries}

\subsection{Polar Codes and Systematic Polar Codes}
In conventional polar code design, $N$ independent copies of a channel $W$ are combined in a recursive manner into a vector channel $W_N$, which is then split into $N$ channels ${W_N^{(i)}, \;1\leq{i}\leq{N}}$, referred to as bit-channels.  The Bhattacharyya parameter $Z(W_N^{(i)})$ is used to identify the quality of bit-channel $i$. A polar code of rate $R{=}\frac{K}{N}$ selects the $K$ most reliable bit-channels (with the smallest $Z(W_N^{(i)})$) to input information bits, and the remaining bit-channel inputs are frozen to zero. We use ${\mathcal A}$ to denote the set of information indices, and ${\mathcal F{=}\mathcal A^c}$ to denote the frozen indices. Let $G{=}F^{\bigotimes n}$ be the $N{\times} N$ matrix that is the $n$-th Kronecker power of $F{=}\left[\begin{array}{cc}1 & 0 \\1 & 1 \\ \end{array} \right]$, where $n{=}\log_2 N$.  The polar encoding process is specified by  $x{=}uG$, where $ x,u\in \mathbb{F}^N, G\in \mathbb{F}^{N\times N} $.  

Ar\i kan showed that a systematic encoder can be realized that maps information bits to positions in the set ${\mathcal B} {=} {\mathcal A}$ in the codeword $x$.  
To be specific, $u_{\mathcal A^c}$ is set to 0,  $x_\mathcal B$ is set to the information vector, and  $u_\mathcal A$ and $x_{\mathcal B^c}$ are found by solving a system of equations~\cite{Arikan2011}.


\subsection{Concatenated Polar Codes}
Our focus in this paper is on concatenated code architectures in which all component codes are polar codes. 
The augmented and flexible length architectures were introduced in~\cite{Elk2017}. In an augmented polar code, a short, rate $R_0=\frac{K_0}{N_0}$  auxiliary outer polar code $G_0$ is connected to an inner polar code $G_1$ of length $N_1$. The $N_0$ bits of the outer codeword are assigned to the semipolarized channels within the inner code (through an interleaver). An additional $K_1$ message bits are assigned to the good bit-channels within the inner code. The total code rate for the augmented structure is $R_{aug} = \frac{K_0+K_1}{N_1}$.

In the flexible length architecture, two inner codes $G_1, G_2$ of length $N_1, N_2$ are coupled through a rate $R_0=\frac{K_0}{N_0}$ auxiliary outer code $G_0$. Information words of length $K_1, K_2$ are assigned to the good bit-channels of the two inner codes, respectively.
The outer codeword  is divided into two parts which are assigned to the semipolarized bit-channels of the inner codes. The total encoding rate for the flexible length structure is $R_{flex} = \frac{K_0+K_1+K_2}{N_1+N_2}$.


Inspired by the flexible length architecture, the local-global polar code architecture,  introduced in~\cite{Zhu2022}, connects multiple inner codes $G_1, ..., G_M$ through a systematic outer polar code. We asssume these codes have the same length $N_i=N, i=1, \ldots, M$. 
A word of $K_b$ information bits is divided into $M$ parts of $K_{b_1}, \ldots, K_{b_M}$ bits that are assigned to good bit-channels within the inner codes.  
The $K_a$ outer information bits are divided into $M$ parts of $K_{a_1}, \ldots, K_{a_M}$ bits that are mapped to semipolarized bit-channels in the $M$ inner codes, respectively. The $P_a$ parity bits  of the outer codeword are similar partitioned into $M$ parts of $P_{a_1}, \ldots, P_{a_M}$ bits and mapped to remaining semipolarized bit-channels within the inner codes. This architecture supports local decoding of information bits $K_{a_i}, K_{b_i}$ within each  inner code $G_i$, with the option of improved decoding of the $M$ inner codewords via global decoding  using the  outer code.    


\subsection{Stopping Set Analysis}
We briefly review the stopping set analysis of polar codes in~\cite{Eslami2013}.
Recall that a stopping set (SS) is a non-empty set of variable nodes such that each neighboring check node is connected to this set at least twice. 
A stopping tree (ST) is a SS that contains one and only one information bit, i.e., variable node in the leftmost stage of the sparse polar code factor graph.  For each information bit $i$, there is a unique stopping tree denoted by $ST(i)$. 
The size of the leaf set (variable nodes on the rightmost stage) of $ST(i)$ is denoted by $f(i)$. Only variable nodes on the right are observed nodes, with all other variables nodes hidden. 
The set of observed variable nodes in a SS  form a variable-node SS (VSS). 
Accordingly, we define a  minimum VSS (MVSS) to be a VSS with a minimum number of observed variable nodes, among all the VSSs. The size of a MVSS is the stopping distance of the code. For any given set $\mathcal J \subseteq \mathcal A$, we denote the set of SSs whose information nodes are precisely $\mathcal J$ as $SS(\mathcal J)$. The set of observed variable nodes in each of these SSs is a VSS for $\mathcal J$, and the collection of these VSSs is   denoted as $VSS(\mathcal{J})$. A minimum size VSS  in $VSS(\mathcal{J})$ is called a minimum VSS for $\mathcal J$, denoted $MVSS(\mathcal{J})$. 


\section{Construction methods}
In this section we describe the stopping set and NDE methods for designing outer codes for augmented and local-global polar codes. 

\subsection{Stopping Set Construction}

\begin{figure}[t]
\centerline{\includegraphics[width=8.5cm,height=5cm]{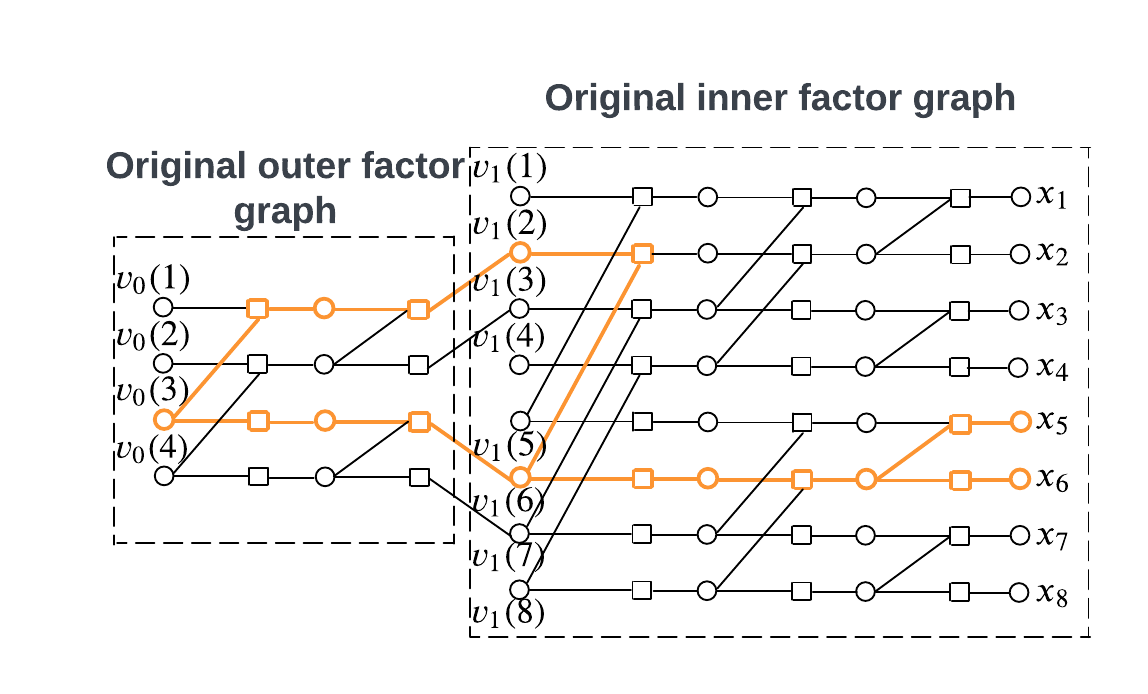}}
\caption{Augmented structure with $N_{0}{=}4$, $N_{1} {=}8$. Orange nodes represent a SS and $\{x_5,x_6\}$ are a MVSS.}
\label{augmented_factor}
\end{figure}

Consider the augmented polar code structure in Fig.~\ref{augmented_factor}, where an inner polar code is concatenated with an outer polar code. Let $v_{0}(i)$ denote the $i$-th node on the leftmost stage within the original outer factor graph, and $v_{1}(j)$ be the $j$-th node on the leftmost stage within the original inner factor graph. Let $\mathcal{J}_i$ denote the set of inner information nodes such that each element in $\mathcal{J}_i$ is connected to one of the leaves in the stopping tree $ST(i)$ defined on the original outer factor graph. For example in Fig.~\ref{augmented_factor}, $\mathcal J_3$ is the set of nodes $\{{v_{1}(2), v_{1}(6)}\}$. Let $MVSS(\mathcal J_i)$ be the MVSS defined by set $\mathcal J_i$ on the original inner factor graph. Let $G_{\mathcal J_i}$ denote the submatrix of the inner encoding matrix $G=F^{\bigotimes n_{1}}$  consisting of the rows that correspond to $\mathcal J_i$. Again taking Fig.~\ref{augmented_factor} as an example, $MVSS(\mathcal J_3)$ is the set of nodes $\{{x_5,x_6}\}$. The resulting $G_{\mathcal J_3}$ consists of the second and sixth rows of inner encoding matrix $G=F^{\bigotimes 3}$. Note that $\{x_1,x_2,x_5,x_6\}$ is also a VSS for $\mathcal J_i$, but it is not the minimum VSS. The following theorem gives a bound on the size of $MVSS(\mathcal J_i)$.

\begin{thm}
Given any outer code position $i$ on the leftmost stage of the factor graph and its corresponding $\mathcal{J}_i$, $1\leq{i}\leq{N_{0}}$, we have $|MVSS(\mathcal J_i)| \geq g(G_{\mathcal J_i})$, where 
\begin{equation}
g(A_{m\times n}) = \sum_{j=1}^{n} \delta(\sum_{i=1}^{m} a_{ij} - 1) 
\end{equation}
\begin{equation}
\delta(x) =
\begin{cases}
    1 & \text{if } x = 0, \\
    0 & \text{otherwise}.
\end{cases}
\end{equation}
\label{th1}
\end{thm}
\renewcommand\qedsymbol{$\blacksquare$}
\begin{proof}
See Appendix.
\end{proof}
In words, the function $g(\cdot)$ counts the number of columns in a matrix that have weight one. Thus, for any given information position $i$ we easily calculate the lower bound on the size of the MVSS that only contains $v_{0}(i)$ on the leftmost stage by looking at the generator matrix of the inner code.

Theorem~\ref{th1} suggests a practical way to design outer polar codes based on the size of stopping sets. We first initialize an unfrozen set $\mathcal O$ for the outer code using the conventional DE algorithm, for example. Then we swap a specified number of  unfrozen bits $i\in{\mathcal O}$ with the smallest ``stopping distance'' $g(G_{\mathcal J_i})$ with some positions $j \in \mathcal{O}^c$ such that $g(G_{\mathcal J_j}) > g(G_{\mathcal J_i})$. 

Before presenting the design algorithm, we introduce some notation.
Let $Q$ be a length $N_0$ vector that contains the indices of bit channels ordered according to channel reliability calculated by DE. The indices are ordered by descending channel reliability, i.e., $Q(1)$ stores the index for the strongest bit channel, $Q(2)$ stores the index for the second strongest, and so on. Let $s$ denote the number of bits we are going to swap. 
For convenience, we denote $g(G_{\mathcal J_i})$ by $g(i)$.
Let $K_0$ denote the size of the desired unfrozen set. Let $min_s(\cdot)$ be the function that returns the $s$-th smallest value in a vector, while $min(\cdot)$ returns the smallest value along with its index. Note that $s$ should be chosen such that there are more than $s$ frozen bits that have $g(\cdot)$ value larger than $min_{s}(g(Q(1)),...,g(Q(K_0)))$. The detailed swapping algorithm is presented in  Algorithm~\ref{alg1}. 

\begin{algorithm}
\caption{Stopping set construction}\label{alg:cap}
\begin{algorithmic}[1]
\Statex \textbf{Input:} $Q$; $g(i)$ for each $i\leq{N_0}$; $s$
\Statex \textbf{Output:} designed unfrozen set $\mathcal O$

\State $threshold = min_{s}(g(Q(1)),...,g(Q(K_0)))$
\State $i \gets 1$
\While{$i \leq s$}
    \State $[value, index] = min(g(Q(1)),...,g(Q(K_0)))$
    \State $j \gets 1$
    \While{True}
        \If{$g(Q(K_0+j)) > threshold$}
            \State $Q(index) \gets Q(K_0+j)$
            \State delete $Q(K_0+j)$ from $Q$
            \State jump to line 14
        \EndIf
        \State $j \gets j+1$
    \EndWhile
    \State $i \gets i+1$
\EndWhile
\State Return $\mathcal{O} = Q(1:K_0)$
\end{algorithmic}
\label{alg1}
\end{algorithm}

We can easily extend Theorem~\ref{th1} to the case when $M$ inner codes are connected by a single outer code. For example, assume $M=2$ and $\mathcal{J}_i = \{ \mathcal{J}_i^1, \mathcal{J}_i^2 \}$, where $\mathcal{J}_i^1$ and $\mathcal{J}_i^2$ are connected nodes in the first and second inner codes, respectively. Then $|MVSS(\mathcal J_i)| \geq ( g(G_{\mathcal J_i^1}) + g(G_{\mathcal J_i^2}) )$. 

The design method of Algorithm~\ref{alg1} can be extended to the local-global polar code architecture, but some care is needed. 
The  systematic outer polar code assigns $M$ information vectors $K_{a_i}$, $i{=}1, \ldots, M$ to  the $M$ inner codes. 
Directly applying Algorithm~\ref{alg1} can
potentially swap bits $K_{a_i}$ with $P_{a_j}$ ($i \neq j$), causing   $[K_{a_i},K_{a_j}]$ to be assigned to the same inner code. For example, assume the unfrozen set $\mathcal{O} = \{ 3,4,7,8\}$ represents the most reliable positions according to DE, and $\mathcal{O}^c = \{ 1,2,5,6\}$. Then, if the partition of $\mathcal{O}$ is according to bit index, the first half of $\mathcal{O}$ will correspond to $K_{a_1}=\{ 3,4\}$ and the second half to  $K_{a_2}=\{ 7,8\}$. If the parity bits are partitioned similarly, we have $P_{a_1}=\{ 1,2\}$ and $P_{a_2}=\{ 5,6\}$. If, after calculating the $G_{\mathcal J_i}$ value  for each position,  we  swap positions 3 and 5, this would yield $K_{a_1}=\{ 4,5\}$. This assignment is now inconsistent with the local-global architecture because part of $K_{a_1}$ (position 5) is connected with the second inner code. To avoid this problem, one needs to carefully choose the connections in the local-global encoder to ensure that positions in $K_{a_i}$ are only swapped with positions in $P_{a_i}$. 

\begin{figure}[htbp]
\centerline{\includegraphics[width=8cm,height=4.2cm]{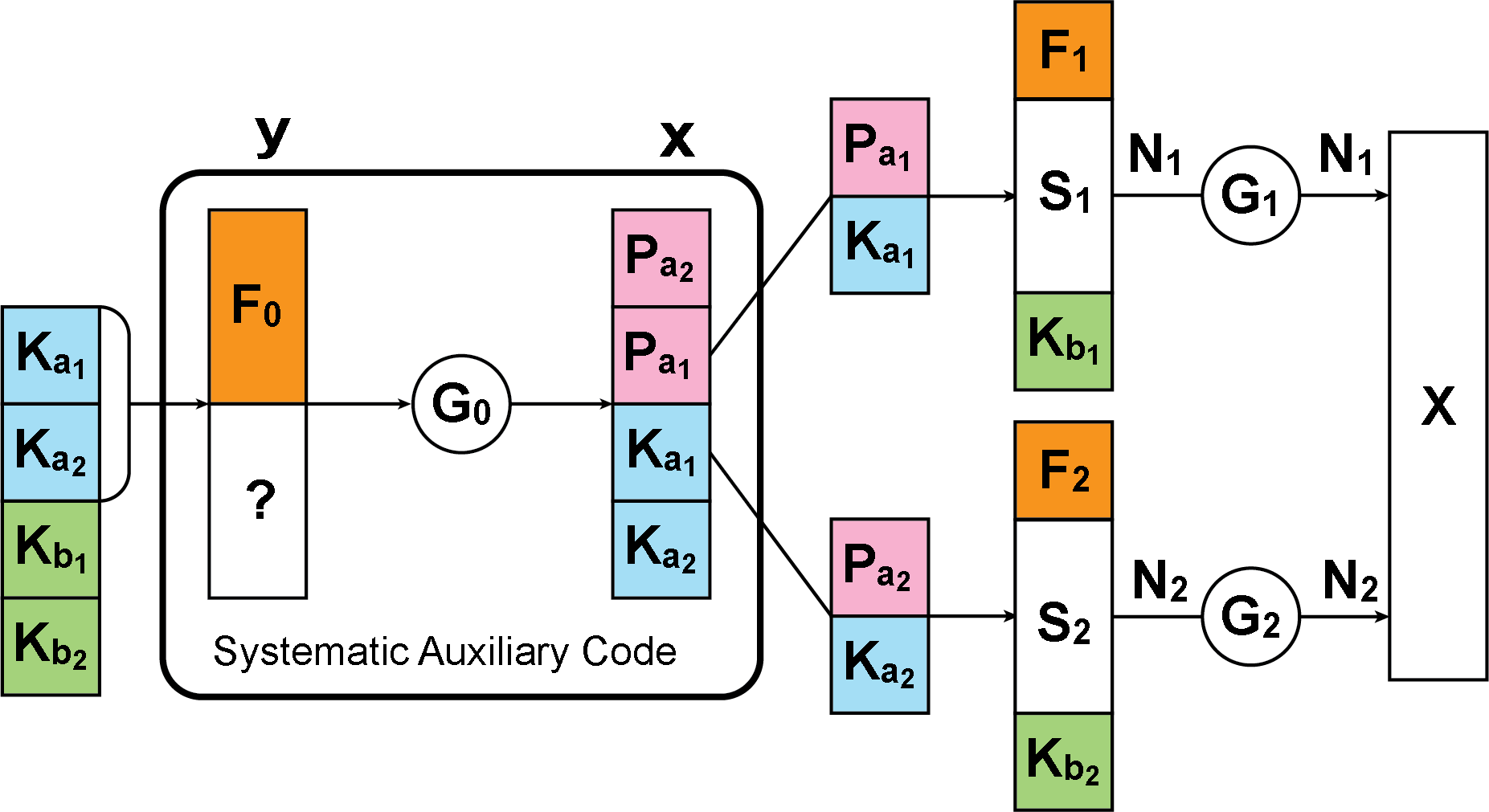}}
\caption{Local-global encoder for stopping set construction.}
\label{Modified_encoder}
\end{figure}

\begin{example}
\label{localglobal}
For the case $M=2$, there is a mapping and partition rule that works for 
any $s < 5$, $2^9 \leq N_1 = N_2 \leq 2^{11}$, $2^6 \leq N_{0} \leq 2^8$ with $R_{0} = R_{all} = \frac{1}{2}$. Let $\mathcal{O}$ be the information set of size $N_{0}/2$ according to DE. Assign to $K_{a_2}$ the first half according to natural index order, and to $K_{a_1}$ the second half. Assign to $P_{a_1}$ the first half of the parity positions in natural index order, and to $P_{a_2}$ the second half. The semipolarized bit-channels in the first inner code are connected in natural index order to  
$K_{a_1}, P_{a_1}$ and those in the second inner code connect to $K_{a_2}, P_{a_2}$. See Fig.~\ref{Modified_encoder}.
\end{example}

\subsection{Nonstationary density evolution (NDE)}

\subsubsection{NDE for augmented polar codes}
In prior formulations of augmented polar codes in~\cite{Guo2014},\cite{Elk2017}, the outer polar codes were designed independently from the code concatenation. 
Since the inputs to the outer code are the LLRs passed from the inner code and the LLR densities may vary within the  semipolarized bit-channels, there is a potential mismatch with the assumed reliability ordering of the outer code bit-channels which could lead to inferior performance of the augmented code.  

In order to improve the performance of the concatenated code under BP decoding, we introduce a small modification of the conventional DE design method introduced in~\cite{DE} which we call the  nonstationary DE (NDE) algorithm. The NDE algorithm does not assume that the inital LLR distributions are identical; rather, the LLR distributions coming from the inner code at the rightmost stage of the outer code factor graph correspond to $N_0$ separate binary symmetric memoryless channels $W_i, i=1, \ldots, N_0$. (A similar nonstationary scenario in code construction via polarization was considered in~\cite{Zor2019}.)

Using the notation of~\cite{DE}, the densities on the rightmost stage are initialized as $\mathrm{a}_{1}^{i} = \mathrm{a}_{W_i}, \;\; 1 \leq i \leq N_{0}$. We then apply DE over the SC decoding trees to compute the LLR densities $\mathrm{a}_{N_0}^i, i=1, \ldots,N_0$ for the bit-channels. 
As in~\cite{DE}, assuming symmetry and  an all-zero codeword, we compute a probability of error, denoted $P(\mathcal{A}_i)$,  for   bit-channel $i$ by integrating  $\mathrm{a}_{N_0}^i$ over the interval $(-\infty, 0)$. The unfrozen set $\mathcal{O}$ is chosen to minimize $\sum_{i\in \mathcal{O}}P(\mathcal{A}_i)$, subject to $|\mathcal{O}|=K_0$. 

In practice, we replace each initial LLR density of the outer code with the empirical  LLR density of the corresponding bit-channel of the inner code after $t$ iterations of BP decoding under assumption of an all-zero codeword. If the semipolarized bit channels of the inner code are denoted by index set $\mathcal{H}$, then the initial LLR density $\mathrm{a}_1^i$ is replaced by the empirical density $b_t^{\mathcal{H}(i)}$ of the corresponding semipolarized bit-channel $\mathcal{H}(i)$.
Details of the NDE  implementation are given in Algorithm~\ref{algo2}.

\begin{algorithm}
\caption{Nonstationary Density Evolution (NDE)}\label{alg:cap}
\begin{algorithmic}
\Statex \textbf{Input:} $\mathcal H$; $K_0$; $b_t^{\mathcal{H}}$
\Statex \textbf{Output:} designed unfrozen set $\mathcal O$
\State $\mathrm{a}_{1}^{i} \gets b_t^{\mathcal H(i)}$, for each $i\leq N_{0}$
\For{column $l$ from right to left}

    \Comment{on original outer factor graph}
    \For{each node in column $l$}
        \State update its density by DE
    \EndFor
\EndFor
\State Return $\mathcal{O}$ which minimizes $\sum_{i\in \mathcal{O}}P(\mathcal{A}_i)$, $|\mathcal{O}|=K_0$

    \Comment{see Theorem 1 in \cite{DE} for $\mathcal{A}_i$ }
\end{algorithmic}
\label{algo2}
\end{algorithm}

\subsubsection{NDE for local-global polar codes}
Let $\mathcal{H}=\{\mathcal{H}_1, \ldots, \mathcal{H}_M\}$ be the positions within the inner codes that are connected to the $N_0$ bits of the outer codeword. The connection mapping requires knowledge of the unfrozen set $\mathcal{O}$ of the outer code. We assume here that the mapping, which we denote by $\pi_{\mathcal{O}}$, respects the natural ordering of bit-channels within the subsets $K_{a_i}$ in $\mathcal{O}$ and $P_{a_i}$ in $\mathcal{O}^c$ and their corresponding subsets within the semipolarized bit-channels of the associated inner codes. Once the connections are established and a set of empirical LLR densities $b_t^{\mathcal{H}}$ are found for the bit-channels in $\mathcal{H}$, we can apply the NDE algorithm to determine a probability of error, denoted $P(\mathcal{A}_i, \pi_{\mathcal{O}}, b_t^{\mathcal{H}})$, for bit-channel $i$ of the outer code.
(All-zero codewords are again assumed in computing the empirical LLRs for the inner code and in computing probabilities of error for the bit-channels of the outer code.) 
We can then determine a set of bit-channels $\mathcal{O'}$ that minimizes $\sum_{i\in\mathcal{O}'} P(\mathcal{A}_i, \pi_{\mathcal{O}}, b_t^{\mathcal{H}})$.

The unfrozen set $\mathcal{O}^*$ that we seek for the outer code is the solution to the equation
 \begin{equation}
\mathop{\arg\min}\limits_{\mathcal{O'}}\sum_{i\in\mathcal{O}'}P(\mathcal{A}_i,\pi_{{\mathcal{O}^*}},b^{\mathcal{H}}_t) = \mathcal{O}^*
\label{opt_unfrozen}
\end{equation}

It is not clear if~(\ref{opt_unfrozen}) has a solution, and, if it does, whether the solution is unique. However, we have used the iterative procedure shown in Algorithm~\ref{search} to find a solution $\mathcal{O}^*$ in the code structures we considered. Different initial values of the unfrozen set can lead to different solutions, and we have also observed cases with periodic solutions, but they  differ in only a small number of bit-channels and the corresponding values of $\sum_{i\in\mathcal{O}^*} P(\mathcal{A}_i, \pi_{{\mathcal{O}^*}}, b_t^{\mathcal{H}})$ are nearly the same. We arbitrarily choose one solution for the outer code design. 


 
\begin{algorithm}
\caption{NDE for local-global structure}\label{alg:cap}
\begin{algorithmic}[1]
\Statex \textbf{Input:} ${\mathcal H}$; $K_0$; $b_t^{\mathcal H}$
\Statex \textbf{Output:} designed unfrozen set $\mathcal O^*$
\State Randomly select $\mathcal{O}^0$, $|\mathcal{O}^0| = K_0$
\For{$m = 1:10$}
    \State Determine $\pi_{\mathcal{O}^{m-1}}$ by $\mathcal{O}^{m-1}$
    \State $\mathrm{a}_{1}^{i}\gets b_t^{\pi_{\mathcal{O}^{m-1}}(i)}$, for each $i\leq N_{0}$
    \For{column $l$ from right to left}
    
        \For{each node in column $l$}
            \State update its density by DE
        \EndFor
    \EndFor
    \State Pick $\mathcal{O}^{m}$ that minimizes $\sum_{i\in \mathcal{O}^{m}}P(\mathcal{A}_i,\pi_{\mathcal{O}^{m-1}},b^{\mathcal{H}}_t)$
    
    \Comment {subject to $|\mathcal{O}^{m}|=K_0$}
    \If{$\mathcal{O}^{m} = \mathcal{O}^{m-1}$}
        \State jump to line 15
    \EndIf
\EndFor
\State Return $\mathcal{O}^* \gets \mathcal{O}^{m}$
\end{algorithmic}
\label{search}
\end{algorithm}

\section{Empirical results}
This section gives empirical results under BP decoding for augmented and local-global polar codes constructed using the proposed methods, along with comparisons to codes designed using conventional DE and LLR evolution methods.
The BP decoding schedules  are the same as those in \cite{Elk2017} for augmented codes and in \cite{Zhu2022} for local-global codes. The maximum number of BP decoder iterations is set at 100. In the SS design, we set the number of bit-channel swaps to $s=4$. In the NDE design for the augmented code, we use $t=3$ BP decoder iterations on the inner code to generate the required empirical LLRs, and for the local-global code, we use $t=4$ iterations. 

\begin{figure}[H] 
\centerline{\includegraphics[width=6.1cm,height=4.6cm]{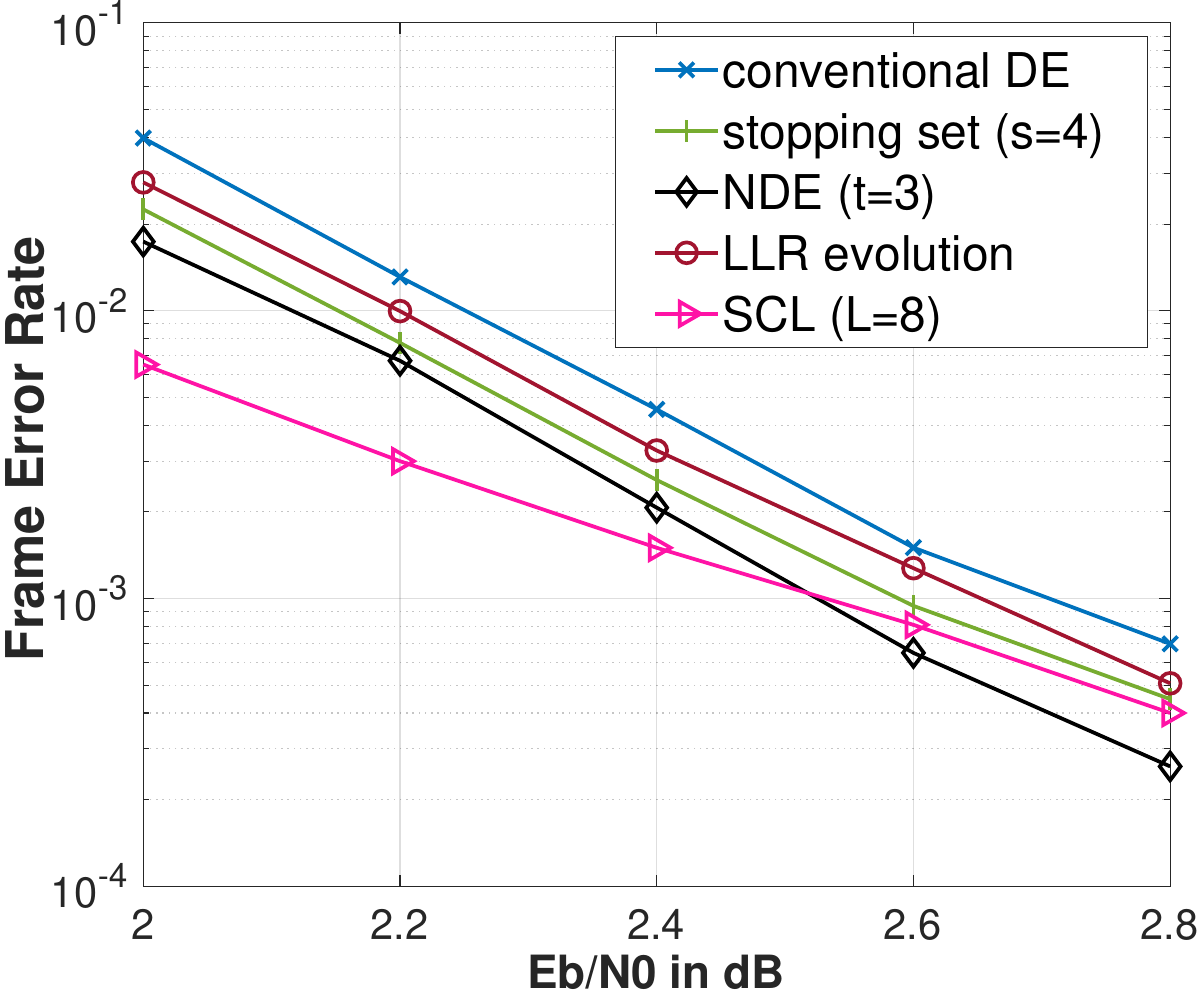}}
\caption{Augmented polar code designs with $N_0{=}64$, $N_1{=}1024$.}
\label{FER_augmented}
\end{figure}

Fig.~\ref{FER_augmented} shows frame error rate (FER) results for augmented code constructions.  The outer code length is $N_0=64$ with code rate $R_0=\frac{1}{2}$. The inner code length is $N_1=1024$. Its bit-channel ordering is based on DE on the AWGN channel at 
$E_b/N_0=3$ dB, computed using the algorithm in \cite{DE}. The design rate of the augmented code is $R_{aug} = \frac{1}{2}$. The connections between the bit-channels of the inner code and the bits of the outer codeword are based on the natural index ordering within the set of  semipolarized bit-channels. 
For the conventional DE and NDE designs, DE is simplified by using Gaussian approximation (GA), using the 4-segment approximation function in~\cite{GA Polar}.

At FER = $10^{-3}$, the stopping set design and NDE design offer gains of  0.12 dB and   0.18 dB over the conventional DE design, respectively, and both outperform the LLR evolution design. 
At this FER they also perform comparably to SCL decoding, although SCL becomes superior at higher FERs. 
We remark that when the outer codes designed using the stopping set and NDE methods are disconnected from the concatenation architecture, their performance is inferior to that of a code designed using conventional DE. This confirms their inherent relationship with the concatenation structure.  



\begin{figure}[H]
\centerline{\includegraphics[width=6.1cm,height=4.6cm]{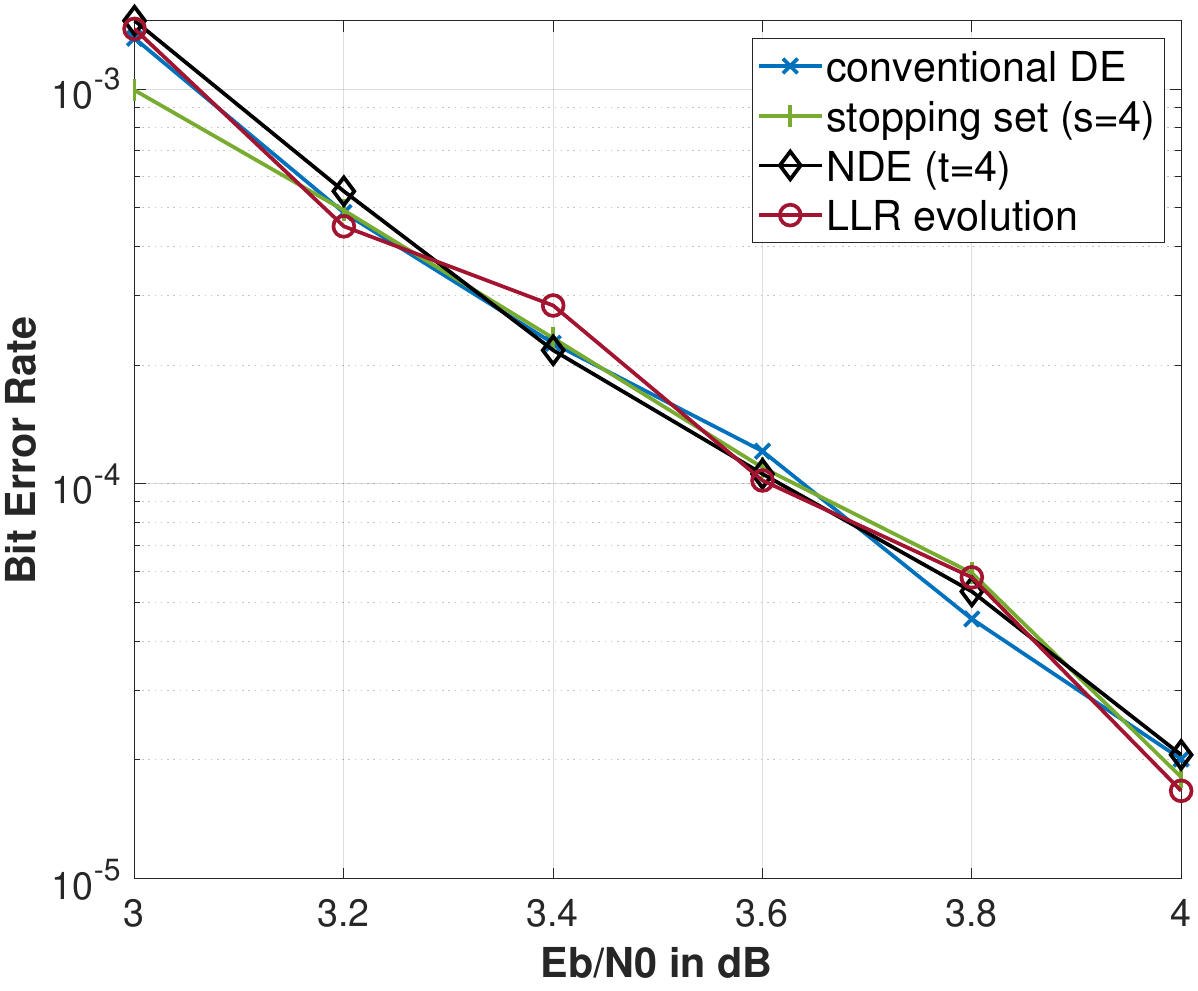}}
\caption{Local decoding results for $N_0{=}256, N_1{=}N_2{=}1024$.}
\label{local}
\end{figure}

\begin{figure}[H]
\centerline{\includegraphics[width=6.1cm,height=4.6cm]{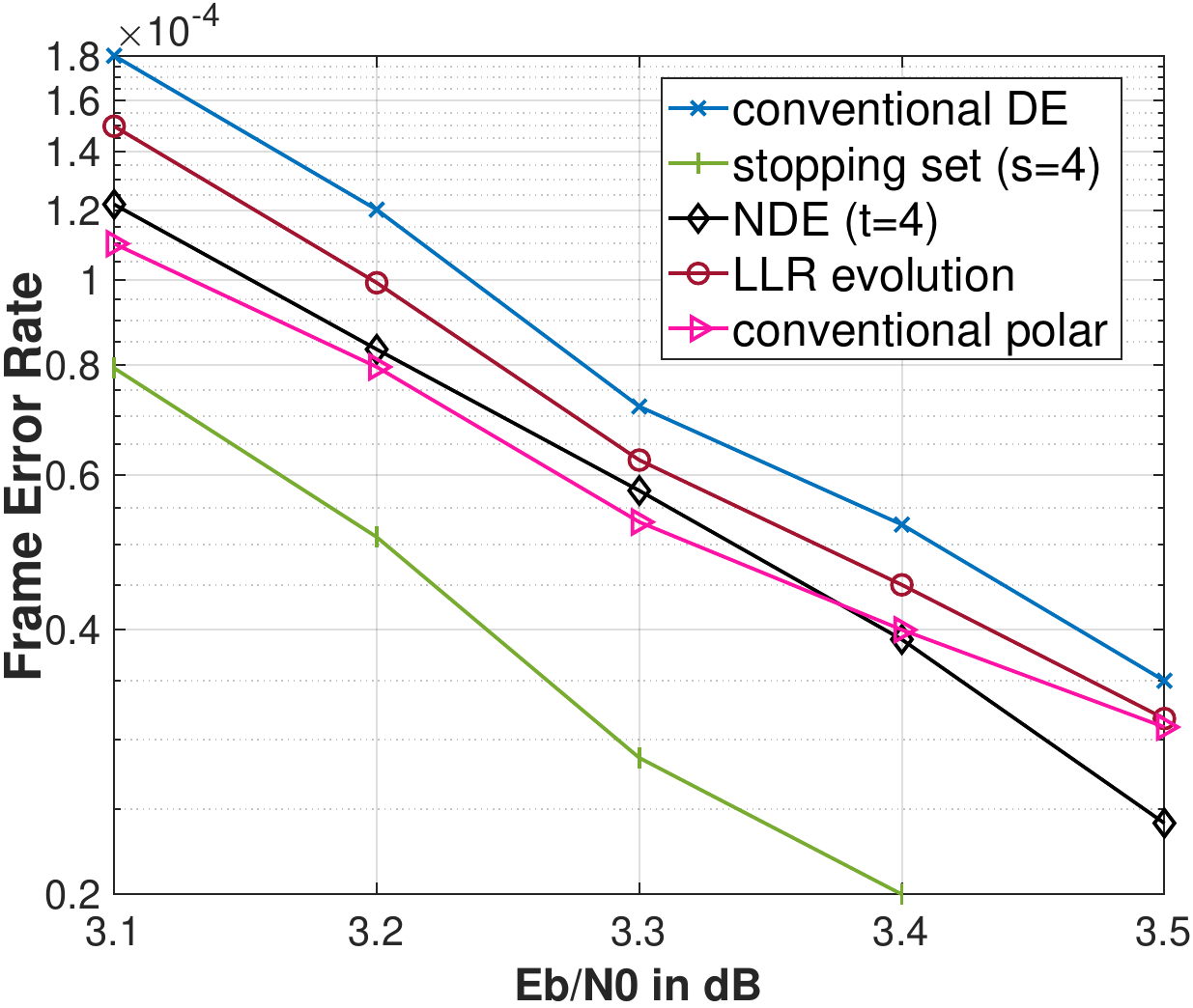}}
\caption{Global decoding results for $N_0{=}256, N_1{=}N_2{=}1024$.}
\label{global}
\end{figure}

Figs.~\ref{local} and~\ref{global} present the results for local and global decoding, respectively, for a local-global code with component code lengths $N_0=256 ,  N_1 = N_2 = 1024$.  The connections between the inner codes and the outer code are as 
described in Example~\ref{localglobal}. Note that Algorithm~\ref{search} for NDE typically converges in fewer than 10 iterations to a unique or oscillating solution. Local decoding results for the different outer code design methods are similar, as expected,  since local decoding does not rely on the outer code. Under global decoding, at FER = $5\times10^{-5}$, the stopping set and NDE designs provide gains of 0.22 dB and 0.09 dB over conventional DE, respectively.  
Both outperform the LLR evolution design.
The result for a length-2048 conventional polar code is also shown for reference.
In summary, the improved global decoding performance provided by the new outer code constructions does not reduce the local decoding performance. 


\section{Conclusion}

In this paper, we proposed two methods to design outer polar codes within concatenated structures. The stopping set and NDE designs both outperform existing construction methods on augmented and local-global architectures with connection patterns based on the natural index ordering of semipolarized bit-channels. Although the natural ordering is attractive from an implementation standpoint, it may not provide the best starting point for these design methods, and experiments with other connection patterns, both structured and randomly generated, show that  gains achieved with the proposed methods vary. Determining the choice of connection pattern and outer code design method that yields the best performance remains a problem for further research. 



\section*{Acknowledgment}
This research was supported in part by NSF Grants CCF-1764104 and CCF-2212437.

\newpage
\;
\newpage

\appendix

\textbf{Proof of Theorem~\ref{th1}}: The definition of stopping set implies that the union of stopping sets is a stopping set.
For a given set of information bits $\mathcal{J}$, consider the set $S(\mathcal{J})$ which is the union  of all the stopping trees defined by the elements in $\mathcal{J}$, i.e., 
$S(\mathcal{J}) = \cup _{j\in \mathcal{J}} ST(j)$.
To find a minimum size VSS for $\mathcal{J}$, we can try to find a stopping set that is properly contained in $S(\mathcal{J})$ that preserves $\mathcal{J}$ but has fewer leaf nodes (i.e., observed variable nodes). We claim that the only leaf nodes that can possibly be deleted from $S(\mathcal{J})$ are shared leaf nodes, i.e., leaf nodes that belong to  at least two stopping trees in $S(\mathcal{J})$.

To see this, suppose that unshared leaf node $v(k,1) \in S(\mathcal{J})$ belongs only to the stopping tree $ST(i)$, where $i\in \mathcal{J}$. The variable nodes on the branch of $ST(i)$ that traces back to the root node $v(i,n+1)$ must also be unshared.\footnote{We are using the indexing convention of~~\cite{DE} for stages in the polar code factor graph. This is the reverse of the indexing convention in~\cite{Eslami2013}.}  This is because if one of the nodes $v(p,q)$ on that branch is shared by two trees or more, then all the children nodes of $v(p,q)$, i.e., nodes to the right of $v(p,q)$ along the tree $ST(i)$,  must be shared nodes, including the leaf node $v(k,1)$. This contradicts the assumption that $v(k,1)$ is unshared.  Since this branch belongs only to $ST(i)$, the result of deleting $v(k,1)$ or any subset of nodes from this branch other than the root node could not produce a stopping set, for this would mean that the remaining subset of nodes in $ST(i)$ would still  constitute a stopping tree, call it $ST'(i)$,
with root $v(i,n+1)$. However, this would violate Fact 2 in~\cite{Eslami2013}, which states that every information bit has a unique stopping tree.

\begin{figure}[htbp]
\centerline{\includegraphics[width=8cm,height=5.5cm]{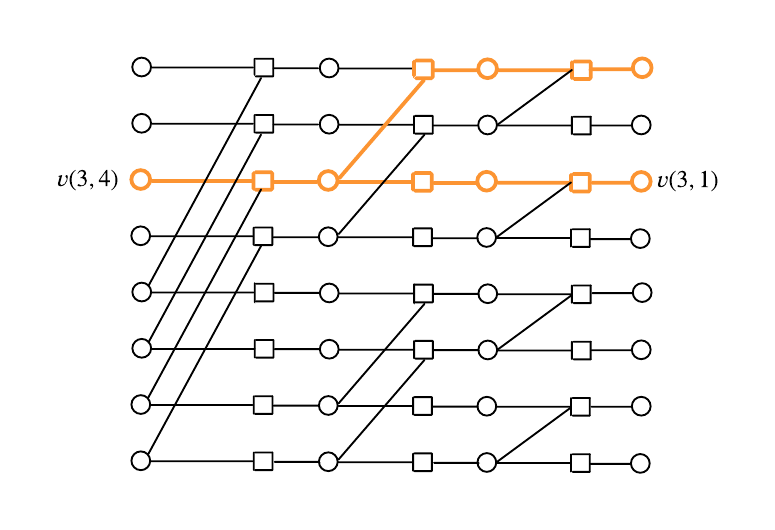}}
\caption{Example of a stopping tree.}
\label{stopping_set}
\end{figure}

A simple example is shown in Fig.~\ref{stopping_set}, where $\mathcal{J}=\{3\}$ and  $S(\mathcal{J})=ST(3)$. There is no stopping set properly contained in  $S(\mathcal{J})$ that does not include  $v(3,1)$.

To complete the proof, we need to characterize the indices of the leaf nodes in $S(\mathcal{J})$. This can be done by noting that the
indices for the leaf nodes in the stopping tree $ST(i)$ are given by the indices for the ones in $r_i^n$, where $r_i^n$ is the $i$-th row of $G^n =F^{\bigotimes n}$, the encoding matrix for length $2^n$ polar codes.  This fact was stated in~\cite{Eslami2013}. For completeness, we provide a detailed proof here. The proof proceeds by induction. For the case $n=1$, the statement follows immediately from inspection of the matrix $G_1=F$ and inspection of the corresponding factor graph for $n=1$.  

\begin{figure}[htbp]
\centerline{\includegraphics[width=8cm,height=6.5cm]{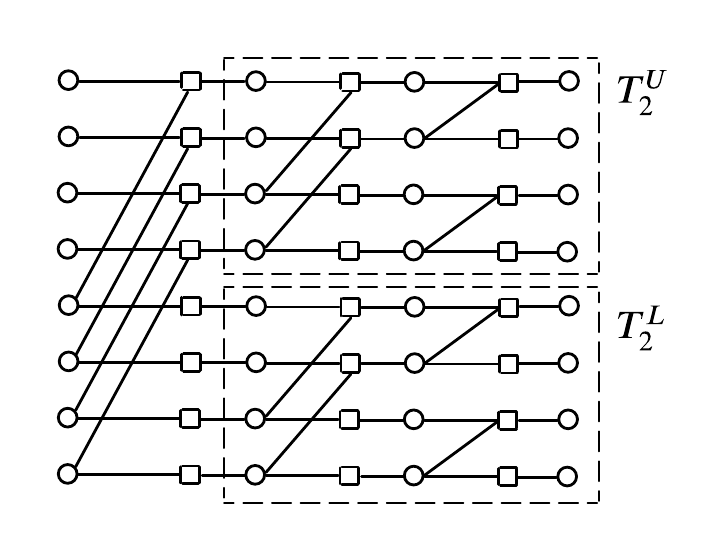}}
\caption{Recursive factor graph structure.}
\label{recursive_graph}
\end{figure}

Now, suppose the result is true for a given $n$. For a length $2^{n+1}$ polar code, we denote the upper and lower halves of the factor graph by $T_n^{U}$ and $T_n^{L}$, as shown in Fig.~\ref{recursive_graph}. The recursive construction of the factor graph implies that $T_n^{U}$ and $T_n^{L}$ are isomorphic, and $G^{n}$ is the encoding matrix for each of the subgraphs $T_n^{U}$ and $T_n^{L}$.  Let $I_i^{n}$ be a length $2^n$ binary vector in which the  indices of ones are the positions of the leaf nodes contained in $ST(i)$ in $T_n^{U}$. By the induction hypothesis, $I_i^{n} = r_i^n$.  

Note that $G^{n+1}$ is recursively represented as 
\begin{equation}
    G^{n+1} = 
            \begin{bmatrix}
                G^n & 0 \\
                G^n & G^n 
            \end{bmatrix}.
\label{recursive_G}
\end{equation}
Referring to Fig.~\ref{recursive_graph}, we see that, for $1\leq i\leq 2^n$,  $I_i^{n+1} = [I_i^n,0,...,0] = [r_i^n,0,...,0] =   r_i^{n+1}$,
with the last equality following from~(\ref{recursive_G}). Similarly, for $2^n+1\leq i\leq 2^{n+1}$, since the subgraphs $T_n^{U}$ and $T_n^{L}$ are isomorphic, we have $I_i^{n+1} = [I_{i-2^n}^n,I_{i-2^n}^n] = [r_{i-2^n}^n,r_{i-2^n}^n]= r_i^{n+1}$, where again the last equality follows from~(\ref{recursive_G}). 
This completes the induction. 

In summary, the number of leaf nodes in $S(\mathcal{J})$ is given by the number of columns in $G_{\mathcal{J}}$ that have non-zero weight. 
The number of unshared leaf nodes in $S(\mathcal{J})$ is given by the number of columns in $G_{\mathcal{J}}$ that have weight exactly one. Thus $g(G_{\mathcal{J}})$ is precisely the number of unshared leaf nodes, which by the previous discussion must belong to a
minimum size VSS for $\mathcal{J}$. This implies $|MVSS(\mathcal{J})|\geq g(G_{\mathcal{J}})$, as desired. 
\qed

The following result shows that the bound in Theorem~\ref{th1} is tight when $|\mathcal{J}|=2$.

\begin{proposition}
When $|\mathcal{J}|=2$, $|MVSS(\mathcal J)| = g(G_{\mathcal J}) $.
\label{a1}
\end{proposition}


\textbf{Proof:} Let $\mathcal{J} = \{i,j\}$. Then $S(\mathcal{J}) = ST(i) \cup ST(j)$. Assume $v(k,1) \in S(\mathcal{J})$ is a shared leaf node belonging to both $ST(i)$ and $ST(j)$.  Then there must exist a degree-3 check node in the graph of $S(\mathcal{J})$, with all three neighboring variable nodes in $S(\mathcal{J})$, that lies in the intersection of the graphs of  $ST(i)$ and $ST(j)$ and has $v(k,1)$ as a child node. Otherwise, one could trace back from $v(k,1)$ to a single root node, contradicting the fact that it is a shared leaf node.  All of the children nodes to the  right of that degree-3 check node must  be shared by the two trees. This implies that we can delete those children nodes, which include the shared node  $v(k,1)$,  and the remaining structure will still be a stopping set in $S(\mathcal{J})$. This procedure can be repeated for any remaining shared leaf nodes, until all of the original shared leaf nodes in $S(\mathcal{J})$ have been deleted. The only remaining leaf nodes are the original unshared leaf nodes, completing the proof. 
\qed

\begin{figure}[htbp]
\centerline{\includegraphics[width=8cm,height=5.5cm]{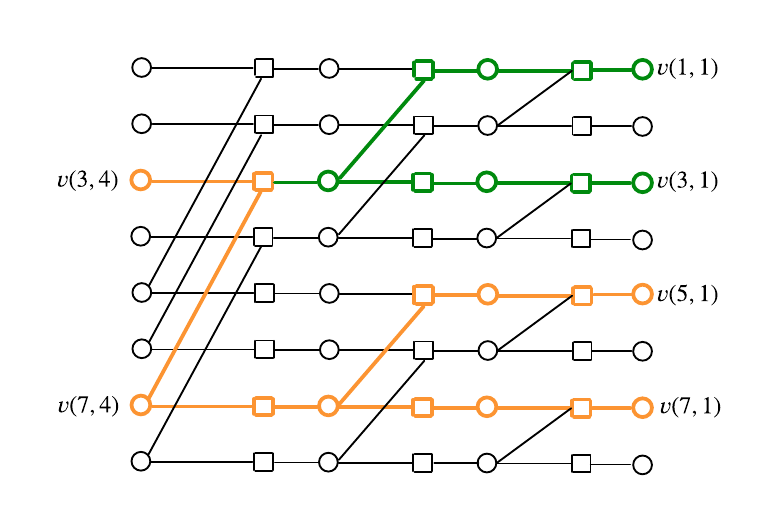}}
\caption{An illustration of Proposition~\ref{a1}.}
\label{MVSS}
\end{figure}

Fig.~\ref{MVSS} illustrates the proof procedure. Here $\mathcal{J} = \{3,7\}$, and $S(\mathcal{J}) = ST(3) \cup ST(7)$. The shared leaf nodes are $v(1,1)$ and $v(3,1)$, while the unshared  leaf nodes are $v(5,1)$ and $v(7,1)$. The green variable nodes in $S(\mathcal{J})$ are the children of the orange degree-3 check node that lies in the  intersection of the graphs of $ST(3)$ and $ST(7)$.  Both $v(1,1)$ and $v(3,1)$ are children variable nodes of this check node. The orange nodes represents a $MVSS(\mathcal{J})$ that remains after deleting the green variable nodes. 


We note that Proposition~\ref{a1} can not be extended to the case when $|\mathcal{J}|>2$. For example, if $\mathcal{J}=\{2,7,8\}$, then the leaf node $v(2,1)$ is shared by $ST(2)$ and $ST(8)$. However, any proper stopping set in $S(\mathcal{J})$ that contains $\mathcal{J}=\{2,7,8\}$ must also contain $v(2,1)$. 

\end{document}